\documentclass[twocolumn,showpacs,preprintnumbers,amsmath,amssymb]{revtex4}
\usepackage{graphicx}
\begin{document}
\title{Instability of Myelin Tubes under Dehydration: deswelling of
  layered cylindrical structures}
\author{C.-M.\ Chen$^1$, C.F.\
  Schmidt$^2$, P.D.\ Olmsted$^3$, and F.C.\ MacKintosh$^{2,4}$}
\affiliation{$^1$Department of Physics, National Taiwan Normal University,
  Taipei, Taiwan, ROC,
$^2$ Division of Physics and
  Astronomy, Vrije University, Amsterdam, Netherlands,
$^3$Department of Physics \& Astronomy and Polymer IRC,
  University of Leeds, Leeds LS2 9JT, UK,
$^4$Michigan Center for Theoretical Physics \& Department of
Physics, University of Michigan, Ann Arbor MI 48109-1120, USA}
\date{\today}
\begin{abstract}
  We report experimental observations of an undulational instability
  of myelin figures. Motivated by this, we examine theoretically the
  deformation and possible instability of concentric, cylindrical,
  multi-lamellar membrane structures. Under conditions of osmotic stress
  (swelling or dehydration), we find a stable, deformed state in which the
  layer deformation is given by $\delta R \propto
  r^{\sqrt{B_A/(hB)}}$, where $B_A$ is the area compression modulus,
  $B$ is the inter-layer compression modulus, and $h$ is the repeat
  distance of layers. Also, above a finite
  threshold of dehydration (or osmotic stress), we find that the
  system becomes unstable to undulations, first with a characteristic
  wavelength of order $\sqrt{\xi d_0}$, where $\xi$ is the
  standard smectic penetration depth and $d_0$ is the thickness of
  dehydrated region.
\end{abstract}
\maketitle
The swelling of bulk phospholipid by water at a temperature higher
than the chain melting temperature can lead to the growth of
multilayer lipid tubules from the interface of bulk lipid and water
\cite{Chapmann,Sakurai1,Sakurai2,Mishima,Buchanan}. These tubules are
of mesoscopic dimensions with a diameter about 20-40 $\mu \mathrm{m}$.
They have the symmetry of a smectic-$A$ liquid crystal and a strong
optical anisotropy. Both electron microscopy \cite{Sakurai3} and X-ray
diffraction show a concentric multilayer structure of these lipid
tubes with a layer spacing about $60 {\rm \AA}$ \cite{Tenchov}. Due to
the similarity of their structural features to those of nerve myelin
sheaths, these tubules are often referred as ``myelin tubes'' or
``myelin figures''. The morphological features of myelin tubes have
been classified into two steps according to the time of growth
\cite{Sakurai2}. During the first step, simple tubes grow into bulk
water with an initial growth rate in length about $1~{\rm \mu m
  /sec}$. The diameter and number of layers of tubes appear to remain
constant during this process and the growth rate is inversely
proportional to the square root of growth time
\cite{Sakurai1,Mishima}. The growth almost stops in the second step
which is characterized by the formation of complicated morphologies of
myelin tubes, such as helical and coiling forms, apparently in order
to maximize their inter-membrane attraction \cite{Mishima2}.

In our experiments, we have observed an instability of myelin
tubes under dehydration by isolating a single myelin tube from
others (during the first step, as shown in Fig.\ 1) and allowing
the bulk water to evaporate in a controlled open chamber. During
this dehydration step, periodic bumps with a wavelength about 1
$\mu {\rm m}$ are observed on the surface of myelin tubes as shown
in Fig.\ 2(a). As water further evaporates, these bumps grow into
arms, as shown in Figs.\ 2(b) and 2(c), and a similar instability
can occur for these long arms. As far as we know, this is the
first reported observation of the instability of myelin tubes
under dehydration. After the bulk water dries out, the tubular
structure of myelins disappear and a mosaic structure is observed.
Although the full structural change of myelin tubes under
dehydration is very complicated and highly nonlinear, the initial
instability of myelin tubes under dehydration poses an interesting
deswelling of layered structures in cylindrical geometry, which we
examine theoretically.

Here, we suggest an explanation for the initial morphological
change of myelin tubes observed in the dehydration step based on
energetic arguments. We first demonstrate and examine a stable
configuration of myelin under conditions of osmotic stress, in
which layer bending does not occur. We then examine the limit of
stability of this to undulations, which relieve in-plane
compression. We predict the appearance of undulations with a
finite wavelength that can be of order micrometers. We then
speculate on the kinetic aspects of the deswelling process. We
note that, unlike the pearling instability \cite{pearling}, the
compression and bending energies already favor finite wavelength
undulations in this case, and the appearance of such an instability
need not be due to hydrodynamic effects. These, we argue will not
significantly change our results for slow drying rates.

For a uni-lamellar tube, dehydration tends to decrease its radius but
increase its length. Surface undulation is disfavored because it
increases tube's bending energy. For a multi-lamellar tube, in
addition to the bending energy of each layer, inter-layer and in-plane
compression energies need to be considered. It can be shown that, in
the limit of long cylinders, a change in length becomes energetically
prohibitive. We show below that undulation of dehydrated layers occurs
only for finite dehydration of a multi-lamellar tube in the presence
of an undisturbed core, while in-plane compression occurs (without
undulations) below this threshold. Although a one-dimensional
undulation along the long axis of myelin tubes is reported in this
paper, we cannot exclude the possibility of a general two-dimensional
undulation along the surface of the tube.

We begin our discussion by considering the dehydration of a myelin
tube in bulk water. The length $L$ of the myelin tube is assumed
to be much larger than its radius $R_0$. As water evaporates, the
osmotic pressure changes due to the increased impurity
concentration in water. In response to the drop in chemical
potential outside, water molecules in the outer layers of the
myelin tube will diffuse out to the bulk and the myelin tube will
deform. This process is relatively fast, as the membranes are
permeable to water. Migration of lipids will be ignored here,
since this is a much slower process due to their low solubility in
water. The removal of water thus results in a radial deformation
$\delta R<0$ of a cylinder of initial radius $R$. Through second
order in this deformation, the area change of this cylinder is
given by $\delta A=2\pi\int dz\left(\delta R+R|\nabla\delta
R|^2/2\right)$. Here, the gradient operator is taken to act in
the plane of the membrane.

We shall find below that there is an initial uniform (in $z$)
deformation $\delta \bar R$ that, beyond a finite threshold,
drives an undulatory instability. We take this undulatory
contribution $\delta R'$ to be sinusoidal. Thus, in order to
determine the onset of this instability, we shall keep terms only
through second order in $\delta R'$. With this in mind, we take
the area strain to be $\epsilon=\delta\bar R/R + |\nabla \delta
R'|^2/2$. The first term in $\epsilon$ is the area deformation
due to the average radius change in a cylinder, while the second
term is due to layer undulation \cite{Helfrich}. This strain is
actually averaged over one period along the cylinder axis. For
fluid membranes, it would not be correct to integrate the square
of the {\em local} area strain, as molecules are free to migrate
within each layer. Rather, the integrated square of $\epsilon$
above properly accounts for the energy associated with area
compression, through second order in $\delta R'$.

We parametrize the structure by the radial position $r$ before
dehydration, and the position $z$ along the axis\cite{2fluid}.
Near the onset of the instability, the displacement field $\delta
R(r,z)=\delta \bar R+\delta R'$ is expected to consist of a
displacement $\delta \bar R(r)$ that is uniform along the cylinder
axis plus a sinusoidally varying displacement $\delta
R'(r,z)\propto\cos \left(qz\right)$. For such a deformation, the
associated energy density in our model can be expressed as
\begin{eqnarray}
f &= & {1\over 2} {B_A \over h} \left({\delta \bar R \over R} +
{1\over 2} |\nabla \delta R'|^2 \right)^2 + {1\over 2} B
\left({\partial
\delta R \over \partial r}\right)^2 \nonumber\\
&& + {1\over 2} K\left(\nabla^2\delta R' \right)^2 ,
\label{denergy}
\end{eqnarray}
where $B_A$ is the in-plane compression modulus, $B$ is the
inter-layer compression modulus, $K$ is the bulk bending
rigidity, and $h$ is the repeat distance of layers before
dehydration.

Before discussing possible undulational instabilities of the
myelin tubes, we first consider deswelling of the structure in
the absence of undulations ({\em i.e.}, $\delta R'=0$). Here, we
ignore the bending contributions. (Note the small curvatures,
with radii of order several micrometers, involved in especially
the outer layers of the myelin, where the deswelling is
greatest.) In this case, the shift in chemical potential of water
molecules in a layer at radius $r^\prime$ due to layer deformation
can be calculated as
\begin{eqnarray}
\mu(r')&-& \mu(r'')
\equiv {\delta F \over \delta N(r')}-{\delta F \over \delta N(r'')}\nonumber\\
 &=& {v_0}\int_{r'}^{r''} \left( {B_A \over h} {\delta
     R \over r^2}
 -{B\over r} {\partial \delta R \over \partial r} -B{\partial^2
 \delta R \over \partial r^2} \right) dr, \label{cpotential}
\end{eqnarray}
where $F\equiv \int(2\pi rf)dr$ is the free energy per unit length,
$N(r^\prime)$ represents the number of water molecules per unit length
in a layer at radius $r^\prime$, and $v_0$ is the molecular volume. In
Eq.\ (\ref{cpotential}), we have used the relation $2\pi r \left[
  \delta R(r)\right] = v_0\delta N(r^\prime) \theta(r-r^\prime)$
resulting from simple geometry, where the Heaviside step function
$\theta(r-r^\prime)$ ensures that $\delta R(r)$ can only be changed by
changing $N$ at $r^\prime < r$. Uniformity of the chemical potential
in equilibrium (at least over time scales for which lipid transport is
negligible) requires the integrand in Eq.\ (\ref{cpotential}) to
vanish. The solution to this differential equation is
\begin{eqnarray}
\delta R \propto r^{\alpha} , \label{eprofile}
\end{eqnarray}
where $\alpha =\sqrt{B_A/(B h)}$. This result applies for both
swelling ($\delta R>0$) and dehydration ($\delta R<0$), where the
amplitude of the displacement field is determined by the value of
the chemical potential outside the cylinder. The result is valid
near the surface where bending contributions are negligible, and
is sensible in that for large $B_A$ there will be little
penetration of the dehydration profile in from the surface, while
for large $B$, the profile is more uniform toward the outside of
the cylinder.

Upon changing the osmotic pressure in the bulk water a dehydration
profile will extend throughout the cylindrical structure. The
initial stage is a smooth deswelling with only a change in the
radius of each layer as calculated above. This also involves
bending energy, as the dehydration profile extends deeper and
becomes more pronounced. For a finite degree of deswelling,
however, the layers can bend to relieve the intra- and inter-
layer compression energies. Hence, we now show that incorporating
the bending degrees of freedom leads to an instability to a
surface undulation. For simplicity, we assume a single mode
approximation and a linear deformation profile that extends to a
depth $d_0$ below the outer surface $R_0$ of the myelin, i.e.,
$\delta R(r=R_0-x)=\left(1-x/d_0\right)\left(\bar a +
  a'\cos(q z)\right)$, where $\bar a$ and $a'$ are the uniform and
undulatory deformation amplitudes, $x$ is the depth below the
outermost layer, and $q$ is the wave number (as shown in Fig.\ 3).
This approximates the state of the system at an instant in time during
dehydration, rather than an equilibrium profile.

In this case, the modulated part of the total free energy per unit
length can then be
calculated from Eq.~(\ref{denergy}) as
\begin{align}
\Delta F_{\rm mod} \simeq \pi d_0 &\left[ \left({K\over
6R_0^3}+{BR_0\over 2d_0^2}\right)\right.\\
&\left.-{B_Aq^2\bar a \over
8h}+{KR_0q^4\over 6}\right] {a'}^{2}. \label{Emodu}
\end{align}
Minimizing Eq.\ (\ref{Emodu}) with respect to the wave number $q$, the
optimal wave number is given by $\left(q^*\right)^2 \simeq {3B_A\bar a
  / 8hKR_0}$. Such an undulation is favored only when $\Delta F_{\rm
  mod}\leq 0$. This results in a threshold
\begin{eqnarray}
\bar a_c=\sqrt{1\over 3}{8hR_0\sqrt{KB}\over B_Ad_0},
\label{Thresh}
\end{eqnarray}
below which the system is stable. The corresponding reduction in
volume of the cylinder is then given by $\Delta V_c=2\pi R_0 L\bar
a_c$. At the onset of instability, we find that the wavelength of
the instability is
\begin{eqnarray}
\lambda \simeq 5 \sqrt{\xi d_0}, \label{lambda}
\end{eqnarray}
which depends only on material parameters and on the depth of the
deswelling profile. Note that any finite range dehydration profile
would lead to a similar result, with different prefactors. For
lipid bilayers, the standard smectic penetration depth
$\xi=\sqrt{K/B}$ is about 0.01 $\mu {\rm m}$.  Therefore the
undulation wavelength is of order 1 $\mu {\rm m}$ if $d_0$ is a
few microns. We have considered a one-dimensional undulation
along the long axis of the myelin tube. Additional undulation
along the $\varphi$ direction might exist.  However, our
calculation shows that such an undulation is disfavored if the
undulation amplitude and wavelength are isotropic in both
directions. We also note that a change in layer spacing due to
surface area compression is much smaller than that due to
dehydration (their ratio is less than $d_0/R_0$) and can be
neglected.

The discussion thus far has been based on energetic
considerations. A dynamic calculation would incorporate the
energetic driving force for the instability and dissipative
damping, and conservation of water in a given layer during the
instability should shift the wavelength $\lambda$ to a smaller
value \cite{two-com}. Moreover, there is the possibility that
large effluence of water induced by a severe dehydration could
induce a dynamic, rather than quasi-equilibrium, instability
through non-linear convective couplings \cite{Catespriv}.

In the early stage of dehydration the dominant process limiting
diffusion is the compression energy between layers, and the
associated diffusion equation for water molecules to cross
bilayers near the surface $r=R_0$ is
\begin{eqnarray}
{\partial \phi \over \partial t}\simeq  B D {\partial^2 \phi
\over \partial r^2}, \label{diffusion}
\end{eqnarray}
where $\phi$ is a coarse-grained composition variable
representing the relative change in water composition (ratio of
interlayer spacing after dehydration to spacing before
dehyeration), and $D$ is a mobility proportional to the membrane
permeability. As dehydration proceeds the intra-layer compression
elasticity $B_A$ will eventually slow this process. In our
calculation above we have approximated the profile $\phi(r,t)$ by
a piecewise linear profile of depth $d_0$, but other bounded
profiles lead to similar results. At a time $t$ after the
external osmotic pressure has been quenched, the penetration of
the dehydration layer will thus scale as $d_0(t) \simeq \sqrt{B D
t}$. Using the piecewise-linear approximation above for $\delta
R$, the total volume change of the outermost cylinder is given by
$\Delta V \cong \pi L R_0 d_0 \left(1- \phi_0 \right)$. The
change in layer spacing of the outermost layer is $\delta R(r) -
\delta R(r-h)=h(\phi_0-1).$

Our instability condition, Eq.~(\ref{Thresh}), assumes a given profile
and dehydration depth.  Assuming further the local equilibrium
assumption, that the chemical potential of the outermost layer is
equal to the chemical potential $\mu_{ext}$ specified by the bulk,
leads to the relation $\mu_{ext}\simeq -v_0 B (1-\phi_0)$.
(Corrections to this involve terms that go as $1/R_0$ and $1/R_0^2$
from Eq.~(\ref{cpotential})). A typical scenario would involve sealing
the chamber after a rapid dehydration to maintain, for a large
water-to-lipid ratio, a constant osmotic pressure $\Pi=\mu_{ext}/v_0$
outside, and hence the relation $(1-\phi_0)\sim\Pi/B$ at the outermost
layer. In this case the critical size of the dehydrated region is
$d_0^c \simeq \sqrt{10 h \xi R_0 B^2/(B_A \Pi)}$. Equivalently, the
critical time scale for the instability to occur after the dehydration
is $\tau \simeq h \xi B R_0/(DB_A\Pi)$, which increases linearly with
$R_0$. The initial undulation wavelength is proportional to
$R_0^{1/4}$.

One could envision another extreme of imposing a constant
dehydration flux $j$ by, for example, exposing the chamber to an
unsaturated vapor that continually increases the impurity
concentration $c_{imp}$ in the bulk water, hence increasing the
osmotic pressure and the driving force for dehydration. In this
case the details of the timescale necessary for the instability
would depend on comparing the rate of change of $\phi_0$ to the
time for diffusion out of the multilamellar state, although the
relation between outer composition and overall volume change,
$\Delta V_c=2\pi R_0 L\bar a_c$, would still hold. For the
outermost layer, the stress of layers must balance the osmotic
pressure difference to the bulk water. In this case $\phi_0(t)\sim
1 - j t$, for a constant flux $ j\sim kT \dot{c}_{imp}/B$ where
$kT$ is the thermal energy, and the critical value of $\phi_0$ is
$\phi_0^c \simeq 1 - 10 B h \xi R_0 /\left(B_A  d_0^2\right)$. The
delay time before undulation is $\tau \simeq \sqrt{h\xi
R_0/\left(B_A D j \right)} \propto \sqrt{R_0/j}$. The initial
undulation wavelength is proportional to
$\left(R_0/j\right)^{1/8}$. At a higher drying rate, the thickness
of the dehydrated region is smaller at the onset of instability
which corresponds to a smaller onset undulation wavelength.
Therefore, in an experiment with controllable humidity, the
undulation wavelength can be varied by controlling the relative
humidity. Nevertheless, we do not have quantitative measurements
yet.

To conclude, we have studied the deformation of myelin tubes under
dehydration by curvature elasticity. The cylindrical geometry is
stable for infinitesimal dehydration, and the surface area of
dehydrated layers is reduced due to an in-plane compression. If an
equilibrium is achieved at this stage, the equilibrium radius
deformation profile can be described by $\delta R \propto
r^{\sqrt{B_A/(hB)}}$. For finite dehydration, the cylindrical
geometry becomes unstable and undulation occurs to reduce the
in-plane compression energy. The critical volume change for this
instability is estimated to be $\Delta V_c \simeq {30 h \xi L
R_0^2 B / (d_0 B_A)}$. The undulation wavelength is predicted to
be about $5 \sqrt{\xi  d_0}$ and is comparable to the observed
undulation wavelength.


This work was supported, in part, by the National Science Council
of Taiwan under grant number NSC 90-2112-M-003 and NSF Grant No.
DMR92-57544. FCM and PDO gratefully acknowledge NATO CRG 960678.
FCM wishes to thank P Pincus and CD Santangelo for discussions of
a related problem.
\begin{figure}
\centerline{\includegraphics[scale=0.4]{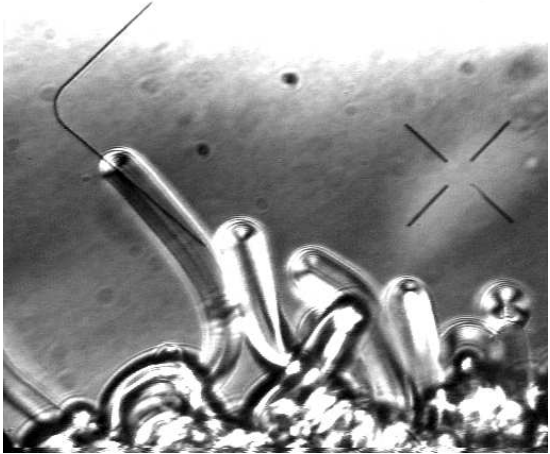}}
\caption{A low resolution view of myelin tubes during the first
  step. The diameter of tubes is about 20-40 microns. An isolated
  myelin tube can be obtained by introducing a water flow across the
  chamber to disintegrate the bulk lipid.} \label{mgrowth}
\end{figure}
\begin{figure}
\centerline{\includegraphics[scale=0.4]{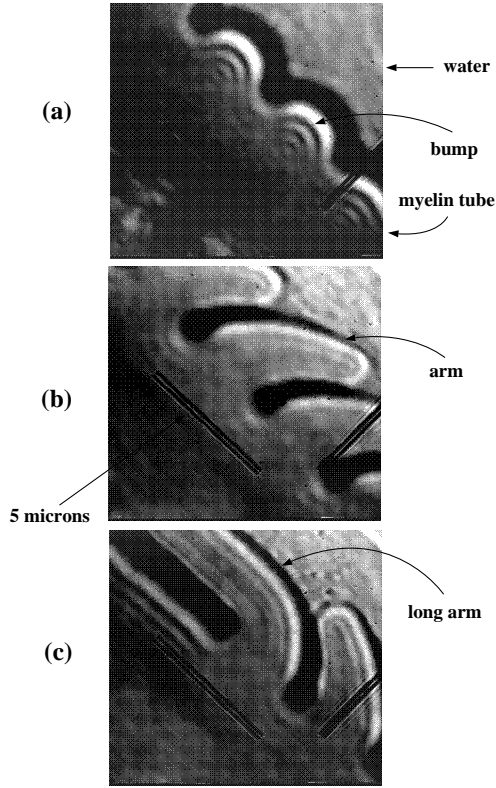}}
\caption{A series of photographs of a myelin tube showing the
  instability under dehydration at different times. The wavelength of
  undulationis about 1 $\mu {\rm m}$. At later times, the amplitude of
  undulation increases and a bump becomes an arm.}
\label{bump}
\end{figure}
\begin{figure}
\includegraphics[scale=0.5]{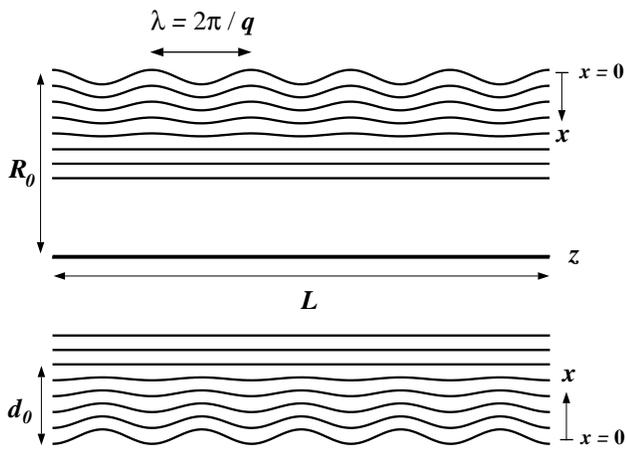}
\caption{Schematic representation of the undulation profile of a
  myelin tube: a cross-sectional view showing different layers.}
\label{undulation}
\end{figure}

\end{document}